\documentstyle[eqsecnum,aps]{revtex}
\newcommand{\hn}{\hat \nabla}

\newcommand{\gab}{g_{ab}}

\newcommand{\gnm}{g^{nm}}

\newcommand{\hTab}{\hat T_{ab}}

\newcommand{\hGab}{\hat G_{ab}}

\begin{document}

\draft

\title{Dual geometries and spacetime singularities}
 
\author{ Israel Quiros\thanks{israel@uclv.etecsa.cu}}
\address{ Departamento de Fisica. Universidad Central de Las Villas. Santa Clara. CP: 54830 Villa Clara. Cuba }

\date{\today}

\maketitle

\begin{abstract}
The notion of geometrical duality is discussed in the context of both Brans-Dicke theory and general relativity. It is shown that, in some particular solutions, the spacetime singularities that arise in usual Riemannian general relativity may be avoided in its dual representation (Weyl-type general relativity). This dual representation provides a singularity-free picture of the World that is physicaly equivalent to the canonical general relativistic one.
\end{abstract}


\section{Introduction}

To our knowledge Dicke was first who raised questions about the physical significance of Riemannian geometry in relativity due to the arbitrariness in the metric tensor resulting from the indefiniteness in the choice of units of measure\cite{bdk,dk}. Actually, Brans-Dicke (BD) theory with a changing dimensionless gravitational coupling constant: $Gm^2\sim\phi^{-1}$ ($m$ is the inertial mass of some elementary particle and $\phi$ is the scalar BD field, $\hbar=c=1$), can be formulated in two equivalent ways for either $m$ or $G$ could vary with position in spacetime\footnotemark\footnotetext{Generaly BD theory can be formulated in an infinite number of equivalent ways since $m$ and $G$ can both vary with position in an infinite number of ways such as to keep $Gm^2\sim\phi^{-1}$.}. The choice $G\sim\phi^{-1}$, $m=const.$, leads to the Jordan frame (JF) BD formalism, that is based on the Lagrangian\cite{bdk}:

\begin{equation}
L^{BD}[g,\phi]=\frac{\sqrt{-g}}{16\pi}(\phi R - \frac{\omega}{\phi} \gnm \nabla_n \phi \nabla_m \phi) + L_{matter}[g],
\end{equation}
where $R$ is the curvature scalar, $\omega$ is the BD coupling constant, and $L_{matter}[g]$ is the Lagrangian density for ordinary matter minimally coupled to the scalar field.

On the other hand, the choice $m\sim\phi^{-\frac{1}{2}}$, $G=const.$, leads to the Einstein frame (EF) BD theory based on the Lagrangian\cite{dk}:

\begin{equation}
L^{BD}[\hat g,\hat \phi]=\frac{\sqrt{-\hat g}}{16\pi}(\hat R - (\omega + \frac{3}{2}) \hat \gnm \hn_n \hat \phi \hn_m \hat \phi) + \hat L_{matter}[\hat g,\hat \phi],
\end{equation}
where now, in the EF metric $\bf{\hat g}$, the ordinary matter is nonminimally couppled to the scalar field $\hat \phi \equiv \ln\phi$ through the Lagrangian density $\hat L_{matter}[\hat g,\hat \phi]$.

Both JF and EF formulations of BD gravity are equivalent representations of the same physical situation\cite{bdk} since they belong to the same conformal class. The EF Lagrangian (1.2) is equivalent to the JF Lagrangian (1.1) in respect to the conformal rescaling of the spacetime metric $\bf g\rightarrow\bf{\hat g}=\phi\bf g$. In the coordinate basis this transformation can be written as:

\begin{equation} 
\hat \gab = \phi \gab,
\end{equation}
where $\phi$ is a nonvanishing smooth function. 

The conformal rescaling (1.3) can be interpreted geometrically as a particular transformation of the physical units (a scalar factor applied to the units of time, length and reciprocal mass)\cite{dk}. Any dimensionless number (for example $Gm^2$) is invariant under (1.3). Experimental observations are unchanged too under these transformations since spacetime coincidences are not affected by them, i.e. spacetime measurements are not sensitive to the conformal rescalings of the metric\footnotemark\footnotetext{Another way of looking at this is realizing that the the experimental measurements deal always with dimensionless numbers and these are unchanged under the transformations of the physical units. For a readable discussion on the dimensionless nature of measurements we recommend section II of reference\cite{am}}. This means that, concerning experimental observations both formulations based on varying $G$ (JFBD) and varying $m$ (EFBD) respectively are indistinguishable. These are physically equivalent representations of a same physical situation.

The same line of reasoning can be applied to the case suggested by Magnano and Sokolowski, involving the conformally related Lagrangians\cite{mag}:

\begin{equation}
L^{GR}[g,\phi]=\frac{\sqrt{-g}}{16\pi}(\phi R - \frac{\omega}{\phi} \gnm \nabla_n \phi \nabla_m \phi) + L_{matter}[g,\phi],
\end{equation}
and

\begin{equation}
L^{GR}[\hat g,\hat \phi]=\frac{\sqrt{-\hat g}}{16\pi}(\hat R - (\omega + \frac{3}{2}) \hat \gnm \hn_n \hat \phi \hn_m \hat \phi) + \hat L_{matter}[\hat g],
\end{equation}
where now, unlike the situation we encountered in usual BD gravity, ordinary matter is minimally coupled in the EF (magnitudes with hat), while it is nonminimally coupled in the JF. Both Lagrangians (1.4) and (1.5) represent equivalent pictures of the same theory: general relativity (GR). Actually, it can be seen that the theory linked with the Lagrangian (1.5) is just GR with a scalar field as an additional source of gravity. In particular, it can be verified that both weak equivalence principle (WEP) and strong equivalence principle (SEP) hold in this case\cite{iq}. We shall call the theory derivable from (1.5) as Einstein frame general relativity(JFGR), while its conformally equivalent representation based on the Lagrangian (1.4) we call Jordan frame general relativity(JFGR).

The field equations derivable from the Lagrangian (1.5) are:

\begin{equation}
\hGab=8\pi\hTab+(\omega+\frac{3}{2})(\hn_a \hat \phi \hn_b \hat \phi- \frac{1}{2} \hat g_{ab} \hat g^{nm} \hn_n \hat \phi \hn_m \hat \phi),
\end{equation}

\begin{equation}
{\hat {\Box}} \hat \phi=0,
\end{equation}
and the conservation equation:

\begin{equation}
\hat \nabla_n \hat T^{na}=0,
\end{equation}
where $\hat G_{ab} \equiv \hat R_{ab}-\frac{1}{2} \hat g_{ab} \hat R$, ${\hat {\Box}}\equiv \hat \gnm \hn_n \hn_m$, and $\hat T_{ab}=\frac{2}{\sqrt{-\hat g}}\frac{\partial}{\partial \hat g^{ab}}(\sqrt{-\hat g}\hat L_{matter})$ are the components of the stress-energy tensor for ordinary matter. 

Now we shall list some features of the JFGR theory that constitute its main disadvantages. The BD scalar field is nonminimally coupled both to scalar curvature and to ordinary matter so the gravitational constant $G$ varies from point to point ($G\sim \phi^{-1}$). At the same time the material test particles don't follow the geodesics of the geometry since they are acted on by both the metric field and the scalar field. This leads that test particles inertial masses vary from point to point in spacetime in such a way as to preserve the constant character of the dimensionless gravitational coupling constant $Gm^2$, i.e. $m \sim \phi^\frac{1}{2}$. The most serious objection to the Jordan frame formulation, however, is associated with the fact that the kinetic energy of the scalar field is not positive definite in this frame. This is usually linked with the formulation  of the theory in unphysical variables \cite{fgn}. In section III we shall show that the indefiniteness in the sign of the energy density in the Jordan frame is only apparent. On the contrary, once the scalar field energy density is positive definite in the Einstein frame it is so in the Jordan one. 

In the present paper we shall focus on those aspects of the Jordan frame formulation of general relativity with an extra scalar field that represent some advantage of this formulation in respect to its conformal EF formulation. It is respecting the transformation properties of the Lagrangian (1.4) under particular transformations of units and the issue of spacetime singularities. In this frame (JF) $R_{mn}k^n k^m$ is negative definite for any non-spacelike vector $\bf k$. This means that the relevant singularity theorems may not hold. This is in contradiction with the Einstein frame formulation of GR where $\hat R_{mn}k^n k^m$ is non-negative and then the occurence of spacetime singularities is inevitable. Then the singularities that can be present in the EFGR may be smoothed out and, in some cases, avoided in the Jordan frame\cite{fgn}.

To the best our knowledge, only the Einstein frame formulation of general relativity (canonical GR and, consequently, Riemann manifolds with singularities it leads) have been paid attention in the literature. This historical omission is the main motivation for the present work.

The paper has been organized as follows. In Sec. II we present the notion of geometrical duality in BD gravity and GR theory. In Sec. III the JF formulation of general relativity is presented in detail. Secs. IV and V are aimed at the study of particular solutions to GR theory that serve as illustrations to the notion of geometrical duality. For simplicity we shall focus mainly in the value $\omega=-\frac{3}{2}$ for the BD coupling constant. In this case EFGR reduces to canonical Einstein's theory\footnotemark\footnotetext{For $\omega=-\frac{3}{2}$ in the EF the scalar field is unphysical and doesn't influence the physics in this frame}. In particular the Schwarzschild solution is studied in Sec. IV, while flat Friedman-Robertson-Walker (FRW) cosmology for perfect fluid ordinary matter with a barotropic equation of state is studied in Sec. V. Finally a physical discussion on the meaning of geometrical duality is given in section VI.

\section{Geometrical duality}

Usually the JF formulation of BD gravity is linked with Riemann geometry\cite{bdk}. It is directly related to the fact that, in the JFBD formalism, ordinary matter is minimally coupled to the scalar BD field through $L_{matter}[g]$ in (1.1). This leads that point particles follow the geodesics of the Riemann geometry. This geometry is based upon the parallel transport law $d\xi^a=-\gamma^a_{mn}\xi^m dx^n$, and the length preservation requirement $dg(\xi,\xi)=0$ where, in the coordinate basis $g(\xi,\xi)=g_{nm}\xi^n\xi^m$, $\gamma^a_{bc}$ are the affine connections of the manifold, and $\xi^a$ are the components of an arbitrary vector $\bf {\xi}$.

The above postulates of parallel transport and legth preservation in Riemann geometry lead that the affine connections of the manifold coincide with the Christoffel symbols of the metric $\bf g$:$\gamma^a_{bc}=\Gamma^a_{bc}=\frac{1}{2}g^{an}(g_{nb,c}+g_{nc,b}-g_{bc,n})$. Under the rescaling (1.3) the above parallel transport law is mapped into:

\begin{equation}
d\xi^a=-\hat \gamma^a_{mn}\xi^m dx^n,
\end{equation}
where $\hat \gamma^a_{bc}=\hat \Gamma^a_{bc}-\frac{1}{2}(\hn_b \hat \phi\delta^a_c+\hn_c \hat \phi\delta^a_b-\hat \nabla^a \hat \phi \hat g_{bc})$ are the affine connections of a Weyl-type manifold given by the length transport law:

\begin{equation}
d \hat g(\xi,\xi)=dx^n \hat \nabla_n \hat \phi \hat g(\xi,\xi).
\end{equation}

In this case the affine connections of the manifold don't coincide with the Christoffel symbols of the metric and, consequently, one can define metric and affine magnitudes and operators on the Weyl-type manifold.

Summing up. Under the rescaling (1.3) Riemann geometry with normal behaviour of the units of measure is mapped into a more general Weyl-type geometry with units of measure varying length in spacetime according to (2.2). At the same time, as shown in section I, JF and EF Lagrangians (of both BD and GR theories) are connected too by the conformal rescaling of the metric (1.3) (together with the scalar field redefinition $\phi \rightarrow \hat \phi= \ln\phi$). This means that, respecting conformal transformation (1.3) JF and EF formulations of the theory on the one hand, and Riemann and Weyl-type geometries on the other, form classes of conformal equivalence. These classes of conformal gravity theories on the one hand, and conformal geometries on the other, can be uniquely linked only after the coupling of the matter fields to the metric has been specified.

In BD theory, for example, matter minimally couples in the JF so the test particles follow the geodesics of the Riemann geometry in this frame, i.e. JFBD theory is naturally linked with Riemann geometry. This means that EFBD theory (conformal to JF one) should be linked with the geometry that is conformal to the Riemann one (the Weyl-type geometry). For general relativity with an extra scalar field just the contrary is true. In this case matter minimally couples in the Einstein frame and then test particles follow the geodesics of the Riemann geometry precisely in this frame, i.e. EFGR is naturally linked with Riemann geometry and, consequently Jordan frame GR (conformal to EFGR) is linked with Weyl-type geometry\footnotemark\footnotetext{When the matter part of the Lagrangian is not present both BD and GR theories can be interpreted on the grounds of either Riemann or Weyl-type geometry indistinctly. We then reach the conclusion that, in this case, both BD theory and general relativity with an extra scalar field coincide. This degeneration of the geometrical interpretation of gravity can be removed only after setting of the matter content of the theory}.

The choice of the unit of length of the geometry is not an experimental issue (for a classical discussion on this subject we refer the reader to \cite{edd}). Moreover, the choice of the spacetime geometry itself is not an experimental issue. We can explain this fact by using a simple argument. The experimental measurements (that always deal with dimensionless numbers) are invariant under the rescaling (1.3) that can be interpreted as a particular units transformation\cite{bdk,dk}. Then physical experiment is insensitive to the rescaling (1.3). The fact that both Riemann and Weyl-type geometries belong to the same equivalence class in respect to the transformation (1.3) completes this explanation.  Actually, this line of reasoning leads that the members in one conformal class are experimentally indistinguishable. 

The same is true for the Jordan frame and Einstein frame formulations of the given classical theory of gravity. The choice of one or another representation for the description of the given physical situation is not an experimental issue. Then a statement such like: 'the JF formulation (or any other formulation) of the given theory (BD or GR theory) is the physical one' is devoid of any physical, i.e. experimentally testable meaning. Such a statement can be taken only as an independent postulate of the theory. This means that the discussion about which conformal frame is the physical one\cite{mag,fgn,yg} is devoid of interest. It is a non-well-posed question. 

An alternative approach can be based on the following postulate. Conformal representations of a given classical theory of gravity are physically equivalent. This postulate leads that the geometrical representation of a given physical situation through general relativity (or BD and Scalar-Tensor(ST) theories in general) produces not just one unique picture of the physical situation but it generates a whole equivalence class of all conformally related pictures. This fact we call as 'geometrical duality'. In this sense Riemann and Weyl-type geometries, for instance, are dual to each other. They provide different geometrical pictures originating from the same physical situation. These different geometrical representations are equally consistent with the observational evidence since they are experimentally indistinguishable. The choice of one or the other picture for the interpretation of the given physical effect is a matter of philosophical prejudice or, may be, mathematical convenience. The word duality is used here in the same context as in \cite{am}, i.e. it has only a semantic meaning and has nothing to do with the notion of duality in string theory.

The rest of this paper is based, precisely, upon the validity of the postulate on the physical equivalence of conformal representations of a given classical theory of gravity. In what follows we shall illustrate the notion of geometrical duality in the context of general relativity with an extra scalar field.

\section{Jordan frame general relativity}

The formulation of general relativity to be developed in the present section is not a complete geometrical theory. Gravitational effects are described here by a scalar field in a Weyl-type manifold, i.e. the gravitational field shows both tensor (spin-2) and scalar (spin-0) modes. In this representation of the theory the redshift effect, for instance, should be interpreted as due in part to a change of the gravitational potential (the metric coefficients) from point to point in spacetime and, in part, as due to a real change in the energy levels of an atom over the manifold\footnotemark\footnotetext{It is due to the fact that, in Jordan frame GR the inertial mass of a given particle varies from point to point in spacetime according to: $m=m_0\phi^\frac{1}{2}$, where $m_0$ is some constant.}.  

The field equations of the Jordan frame GR theory can be derived, either directly from the Lagrangian (1.4) by taking its variational derivatives respect to the dynamical variables or by conformally mapping eqs.(1.6-1.8) back to the JF metric according to (1.3), to obtain:

\begin{equation}
G_{ab}=\frac{8\pi}{\phi} T_{ab}+\frac{\omega}{\phi^2}(\nabla_a  \phi \nabla_b \phi- \frac{1}{2} g_{ab} g^{nm} \nabla_n \phi \nabla_m \phi)+\frac{1}{\phi}(\nabla_a \nabla_b \phi-g_{ab} \Box \phi),
\end{equation}
and

\begin{equation}
\Box \phi=0,
\end{equation}
where $T_{ab}=\frac{2}{\sqrt{-g}}\frac{\partial}{\partial g^{ab}}(\sqrt{-g} L_{matter})$ is the stress-energy tensor for ordinary matter in the Jordan frame. The energy is not conserved because the scalar field $\phi$ exchanges energy with the metric and with the matter fields. The corresponding dynamic equation is:

\begin{equation}
\nabla_n T^{na}=\frac{1}{2} \phi^{-1} \nabla^a \phi T,
\end{equation}

The equation of motion of an uncharged, spinless mass point that is acted on by both the JF metric field $\bf g$ and the scalar field $\phi$,

\begin{equation}
\frac{d^2x^a}{ds^2}=-\Gamma^a_{mn} \frac{dx^m}{ds} \frac{dx^n}{ds}-\frac{1}{2} \phi^{-1} \nabla_n \phi(\frac{dx^n}{ds}\frac{dx^a}{ds}-g^{an}),
\end{equation}
does not coincide with the geodesic equation of the JF metric. This (together with the more complex structure of the equation (3.1) for the metric field in respect to the corresponding equation (1.6)) introduces additional complications in the dynamics of the matter fields.

We shall point out that the different solutions to the wave equation (3.2) generate different Weyl-type geometrical pictures that are dual to the Einstein frame one.

One of the most salient features of the Jordan frame GR theory is that, in general, the energy conditions do not hold due, on the one hand to the term with the second covariant derivative of the scalar field in the righthand side (r.h.s.) of eq.(3.1) and, on the other to the constant factor in the second term that can take negative values. This way the r.h.s. of eq.(3.1) may be negative definite leading that some singularity theorems may not hold and, as a consequence, spacetime singularities that can be present in canonical Riemannian GR (given by eqs.(1.5-1.8)), in Weyl-type GR (JFGR) spacetimes may become avoidable. 

In the following sections we shall illustrate this feature of GR theory in some typical situations where the BD coupling constant is taken to be $\omega=-\frac{3}{2}$.\footnotemark\footnotetext{In BD gravity the case with $\omega=-\frac{3}{2}$ can't be studied because it leads that the field equations of the theory are undefined.} In this case, in the EF the scalar field stress-energy tensor ($\frac{\phi}{8\pi}$ times the second term in the righthand side of eq.(1.6)) vanishes so we recover the canonical Einstein's GR theory with ordinary matter as the only source of gravity \footnotemark\footnotetext{Usual Einstein's GR theory can be approached as well if we set $\phi=const.$.}. Then the EF scalar field $\hat \phi$  (fulfilling the field equation (1.7)) is a non interacting (nor with matter nor with curvature), massless, uncharged, and spinless, 'ghost' field (it is an unphysical field). Nevertheless it influences the physics in the JF. Then its functional form in the EF must be taken into account. For $\omega > -\frac{3}{2}$, $\hat \phi$ is a physical field in the EF.

The fact that the Jordan frame formulation does not lead to a well defined energy-momentum tensor for the scalar field is the most serious objection to this representation of the theory\cite{fgn}. For this reason we shall briefly discuss on this. The kinetic energy of the JF scalar field is negative definite or indefinite unlike the Einstein frame where for $\omega > -\frac{3}{2}$ it is positive definite. This implies that the theory does not have a stable ground state (that is necessary for a viable theory of classical gravity) implying that it is formulated in unphysical variables\cite{fgn}.

We shall point out that, although in this frame the r.h.s. of eq.(3.1) does not have a definite sign (implying that some singularity theorems may not hold), the scalar field stress-energy tensor can be given the canonical form. In fact, as pointed out in reference \cite{ss}, the terms with the second covariant derivatives of the scalar field contain the connection, and hence a part of the dynamical description of gravity. For instance, a new connection was presented in \cite{ss} that leads to a canonical form of the scalar field stress-energy tensor in the JF. 

We can obtain the same result as in ref.\cite{ss} if we rewrite equation (3.1) in terms of affine magnitudes in the Weyl-type manifold (see section II). In this case the affine connections of the JF (Weyl-type) manifold $\gamma^a_{bc}$ are related with the Christoffel symbols of the JF metric through: $\gamma^a_{bc}=\Gamma^a_{bc}+\frac{1}{2} \phi^{-1}(\nabla_b \phi\delta^a_c+\nabla_c \phi\delta^a_b-\nabla^a \phi g_{bc})$. We can define the 'affine' Einstein tensor $^\gamma G_{ab}$ given in terms of the affine connections of the manifold $\gamma^a_{bc}$ instead of the Christoffel symbols of the Jordan frame metric $\Gamma^a_{bc}$. Equation (3.1) can then be rewritten as:

\begin{equation}
^\gamma G_{ab}=\frac{8\pi}{\phi} T_{ab}+\frac{(\omega+\frac{3}{2})}{\phi^2}(\nabla_a  \phi \nabla_b \phi- \frac{1}{2} g_{ab} g^{nm} \nabla_n \phi \nabla_m \phi),
\end{equation} 
where now $\frac{\phi}{8\pi}$ times the second term in the r.h.s. of this equation has the canonical form for the scalar field stress-energy tensor. We shall call this as the 'true' stress-energy tensor for $\phi$, while $\frac{\phi}{8\pi}$ times the sum of the 2nd and 3rd terms in the r.h.s. of eq.(3.1) we call as the 'effective' stress-energy tensor for the BD scalar field $\phi$. Then once the scalar field energy density is positive definite in the Einstein frame it is so in the Jordan frame. This way the main physical objection against this formulation of general relativity is removed. 

Another remarkable feature of the Jordan frame GR theory is that it is invariant in form under the following conformal transformations (as pointed out in ref.\cite{bdk,dk} these can be interpreted as transformations of physical units):

\begin{equation}
\tilde g_{ab}=\phi^2 g_{ab},
\end{equation}

\begin{equation}
\tilde \phi=\phi^{-1},
\end{equation}
and

\begin{equation}
{\tilde g_{ab}}=f g_{ab},
\end{equation}

\begin{equation}
\tilde \phi=f^{-1}\phi,
\end{equation}
where $f$ is some smooth function given on the manifold. In both cases the invariance in form of the equations (3.1-3.4) can be verified by direct substitution of (3.6) and (3.7) or (3.8) and (3.9) in these equations. Also Jordan frame GR based on the Lagrangian (1.4) is invariant in respect to the more general rescaling\cite{iq} (first presented in \cite{far}):

\begin{equation}
\tilde \gab=\phi^{2\alpha}\gab,
\end{equation}
and the scalar field redefinition:

\begin{equation}
\tilde \phi=\phi^{1-2\alpha}.
\end{equation}

This transformation is accompanied by a redefinition of the BD coupling constant:

\begin{equation}
\tilde \omega=\frac{\omega-6\alpha(\alpha-1)}{(1-2\alpha)^2},
\end{equation}
with $\alpha \neq \frac{1}{2}$. The case $\alpha = \frac{1}{2}$ constitute a singularity in the transformations (3.10-3.12).\footnotemark\footnotetext{It should be stressed that the full Jordan frame general relativity theory is invariant in respect to transformations (3.10-3.12). Unlike this  in Jordan frame Brans-Dicke gravity the presence of ordinary matter with $T\equiv T^n_n \neq 0$ breaks this symmetry\cite{far}.} 

The conformal invariance of a given theory of gravitation (i.e. its invariance under a particular transformation of physical units) is a very desirable feature of any classical gravity theory that would correctly describe our real world in the large scale. As it was pointed out by Dicke\cite{dk}, it is obvious that the particular values of the units of mass, length and time employed are arbitrary so the physical laws must be invariant under these transformations. This simple argument suggests that the Jordan frame formulation of general relativity with an extra scalar field that is based on the Lagrangian (1.4) is a better candidate for such ultimate classical theory of gravitation than the other classical theories that are given by the Lagrangians (1.1), (1.2) and (1.5) respectively. In fact, the Lagrangian (1.4) is invariant in respect to the particular transformations of the physical units studied here ((3.6,3.7), (3.8,3.9) and (3.10-3.12)) while the Lagrangians (1.1), (1.2) and (1.5) are not invariant under these transformations. 

In the following section we shall discuss on geometrical duality among singular Schwarzschild (EF) vacuum solution and the corresponding non singular JF solution and, in section V, we shall illustrate this kind of duality for flat, perfect fluid Friedman - Robertson - Walker (FRW) cosmologies. A similar discussion on conformal transformations between singular and non singular spacetimes in the low-energy limit of string theory can be found in \cite{stw} for axion - dilaton black hole solutions in $D=4$ and in \cite{cew} for classical FRW axion - dilaton cosmologies\footnotemark\footnotetext{References \cite{stw,cew} were pointed out to us by D. Wands}. For spurious black hole in the classical approximation see \cite{fik}.

\section{Geometrical duality and Schwarzschild black hole}

In this section, for simplicity, we shall interested in the static, spherically symmetric solution to Riemannian general relativity (Einstein frame GR) with $\omega=-\frac{3}{2}$ for material vacuum, and in its dual Weyl-type picture (Jordan frame GR). In the EF the field equations (1.6-1.8) can be written, in this case, as:

\begin{eqnarray}
\hat R_{ab}=0, \nonumber\\
{\hat {\Box}} \hat \phi=0.
\end{eqnarray}

The corresponding solution , in Schwarzschild coordinates,looks like($d\Omega^2=d\theta^2+\sin^2 \theta d\varphi^2$):

\begin{equation}
d\hat s^2=-(1-\frac{2m}{r}) dt^2+ (1-\frac{2m}{r})^{-1} dr^2+ r^2 d\Omega^2,
\end{equation}
and

\begin{equation}
\hat \phi=q \ln (1-\frac{2m}{r}),
\end{equation}
where $m$ is the mass of the point, located at the coordinate beginning, that generates the gravitational field and $q$ is an arbitrary real parameter. As seen from eq.(4.2) the static, spherically symmetric solution to eq.(4.1) is just the typical Schwarzschild black hole solution for vacuum. The corresponding solution for JFGR can be found with the help of the conformal rescaling of the metric (1.3) and the scalar field redefinition $\phi=e^{\hat \phi}=(1-\frac{2m}{r})^q$:

\begin{equation}
ds^2=-(1-\frac{2m}{r})^{1-q} dt^2+(1-\frac{2m}{r})^{-1-q} dr^2+\rho^2d\Omega^2,
\end{equation}
where we have defined the proper radial coordinate $\rho=r(1-\frac{2m}{r})^{-\frac{q}{2}}$. In this case the curvature scalar is given by: 

\begin{equation}
R=-\frac{3}{2} \phi^{-2} g^{nm} \nabla_n \phi \nabla_m \phi=-\frac{6 m^2 q^2}{r^4}(1-\frac{2m}{r})^{q-1}.
\end{equation}

The real parameter $q$ labels different spacetimes ($M,g^{(q)}_{ab},\phi^{(q)}$), so we obtained a class of spacetimes \{$(M,g^{(q)}_{ab},\phi^{(q)})/q\in \Re$\} that belong to a bigger class of known solutions\cite{alac}. These known solutions are given, however, for an arbitrary value of the coupling constant $\omega$.

We shall outline the more relevant features of the solution given by (4.4). For the range $-\infty<q<1$ the Ricci curvature scalar (4.5) shows a curvature singularity at $r=2m$. For $-\infty<q<0$ this represents a timelike, naked singularity at the origin of the proper radial coordinate $\rho=0$. We shall drop these spacetimes for they are not compatible with the cosmic censorship conjecture\cite{rp}. Situation with $q=0$ is trivial. In this case the conformal transformation (1.1) coincides with the identity transformation that leaves the theory in the same frame. For $q>0$, the limiting surface $r=2m$ has the topology of an spatial infinity so, in this case, we obtain a class of spacetimes with two asymptotic spatial infinities\footnotemark\footnotetext{For $0<q<1$ the spatial infinity at $r=\infty$ is Ricci flat, meanwhile, the one at $r=2m$ is singular. When $q\geq1$ both spatial infinities are Ricci flat} one at $r=\infty$ and the other at $r=2m$, joined by a wormhole with a throat radius $r= (2+q)m$, or the invariant surface determined by $\rho_{min} =q(1+\frac{2}{q})^{1+\frac{q}{2}} m$. The wormhole is asymmetric under the interchange of the two asymptotic regions ($r=\infty$ and $r=2m$)\cite{vh}. 

This way, Weyl-type spacetimes dual to the Riemannian Schwarzschild black hole one (line element (4.2)) are given by the class \{$(M,g^{(q)}_{ab},\phi^{(q)})/q>0$\} of wormhole (singularity free)spacetimes. 

Although in the present paper we are interested in the particular value $\omega=-\frac{3}{2}$ of the BD coupling constant, it will interesting however to discuss, briefly, what happen for $\omega>-\frac{3}{2}$. In this case there is a physical scalar in the Einstein frame (see eq.(1.6)). The corresponding EF solution to eqs.(1.6) and (1.7) is given by\cite{alac}: 

\begin{equation}
d\hat s^2=-(1-\frac{2m}{pr})^p dt^2+ (1-\frac{2m}{pr})^{-p} dr^2+\hat \rho^2 d\Omega^2,
\end{equation}
and

\begin{equation}
\hat \phi=q \ln (1-\frac{2m}{pr}),
\end{equation}
where $p^2+(2\omega+3)q^2=1$, $p>0$. For non-exotic scalar matter ($\omega \geq -\frac{3}{2}$), $0<p\leq 1$. In eq.(4.6) we have used the definition $\hat \rho=(1-\frac{2m}{pr})^\frac{1-p}{2}r$ for the EF proper radial coordinate. There is a time-like curvature singularity at $r=\frac{2m}{p}$, so the horizon is shrunk to a point. Then in the EF the validity of the cosmic censorship hypothesis and, correspondingly, the ocurrence of a black hole are uncertain\cite{alac}.

The JF solution conformally equivalent to (4.6) is given by:

\begin{equation}
ds^2=-(1-\frac{2m}{pr})^{p-q} dt^2+ (1-\frac{2m}{pr})^{-p-q} dr^2+ \rho^2 d\Omega^2,
\end{equation}
where the JF proper radial coordinate $\rho=r(1-\frac{2m}{pr})^{\frac{1-p-q}{2}}$ was used. In this case, when $\omega$ is in the range $0 < \omega+3 < \frac{1+p}{2(1-p)}$, the Weyl-type JF geometry shows again two asymptotic spatial infinities joined by a wormhole. The particular value $p=1$ corresponds to the case of interest $\omega=-\frac{3}{2}$.

The singularity-free character of the Weyl-type geometry should be tested with the help of a test particle that is acted on by the JF metric in eq.(4.8) and by the scalar field $\phi=(1-\frac{2m}{pr})^q$. Consider the radial motion of a time-like test particle($d\Omega^2=0$). In this case the time-component of the motion equation (3.4) can be integrated to give:

\begin{equation}
\dot t^2=-C_1^2(1-\frac{2m}{pr})^{q-2p}, 
\end{equation}
where $C_1^2$ is some integration constant and the overdot means derivative with respect to the JF proper time $\tau$($d\tau^2=-ds^2$). The integration constant can be obtained with the help of the following initial conditions: $r(0)=r_0$, $\dot r(0)=0$, meaning that the test particle moves from rest at $r=r_0$. We obtain $C_1^2=-(1-\frac{2m}{pr_0})^p$. Then the proper time taken for the particle to go from $r=r_0$ to the point with the Schwarzschild radial coordinate $r_0\leq r < \frac{2m}{p}$ is given by:

\begin{equation}
\tau=\int_r^{r_0}\frac{r^{\frac{q}{2}} dr}{\sqrt{(1-\frac{2m}{pr_0})^p-(1-\frac{2m}{pr})^p}(r-\frac{2m}{p})}.  
\end{equation}

While deriving this equation we have used eq.(4.8) written as:$-1=(1-\frac{2m}{pr})^{p-q} d\dot t^2- (1-\frac{2m}{pr})^{-p-q} d\dot r^2$. The integral in the r.h.s. of eq.(4.10) can be evaluated to obtain($q\neq 2$):

\begin{equation}
\tau>\frac{(\frac{2m}{p})^{\frac{q}{2}}(1-\frac{2m}{pr_0})^{\frac{p}{2}}}{1-\frac{q}{2}}[(r_0-\frac{2m}{p})^{1-\frac{q}{2}}-(r-\frac{2m}{p})^{1-\frac{q}{2}}], 
\end{equation}
and

\begin{equation}
\tau>\frac{2m}{p} \ln(\frac{r_0-\frac{2m}{p}}{r-\frac{2m}{p}}),  
\end{equation}
for $q=2$. For $q\geq 2$ the proper time taken by the test particle to go from $r=r_0$ to $r=\frac{2m}{p}$ is infinite showing that the particle can never reach this surface (the second spatial infinity of the wormhole). Then the time-like test particle does not see any singularity.

If we consider the scalar field $\phi$ as a perfect fluid then we find that its 'true' energy density (the (0,0) component of $\frac{\phi}{8\pi}$ times the second term in the r.h.s. of eq.(3.5)) as measured by a comoving observer is given by:

\begin{equation}
\mu^\phi=\frac{2m^2 q^2 (\omega+\frac{3}{2})}{8\pi p^2 r^4}(1-\frac{2m}{pr})^{p+2q-2},  
\end{equation}
while its 'effective' energy density (the (0,0) component of $\frac{\phi}{8\pi}$ times the sum of the second and third terms in the r.h.s. of eq.(3.1)) is found to be:

\begin{equation}
\mu^\phi_{eff}=\frac{2m^2 q(q(\omega+1)-p)}{8\pi p^2 r^4}(1-\frac{2m}{pr})^{p+2q-2}.  
\end{equation}

These are everywhere non-singular for $q\geq\frac{2-p}{2}$ ($0<p\leq 1$) in the range $2m\leq r<\infty$. The 'true' BD scalar field energy density $\mu^\phi$ is everywhere positive definite for $\omega>-\frac{3}{2}$ for all $q$ and $0<p\leq 1$. This means that the scalar matter is non-exhotic and shows a non-singular behaviour evrywhere in the given range of the parameters involved. The scalar field 'effective' energy density $\mu^\phi_{eff}$ is everywhere positive definite only for $q>\frac{p}{\omega+1}$. 
 
Summing up. With the help of time-like test particles that are acted on by both the metric field and the scalar field we can test the absence of singularities (and black holes) in Weyl-type spacetimes of the class \{$M, g_{ab}^{(q)}, \phi^{(q)}/ q\geq 2$\}. These are dual to Riemannian (singular) spacetimes ($M, \hat g_{ab}$) given by (4.2). Pictures with and without singularity are different, but physically equivalent (dual) geometrical representations of the same physical situation. Experimental evidence on the existence of a black hole (enclosing a singularity), obtained when experimental data is interpreted on the grounds of Riemann geometry (naturally linked with Einstein frame GR theory with $\omega=-\frac{3}{2}$) can serve, at the same time, for evidence on the existence of a wormhole when the same experimental data is interpreted on the grounds of the Weyl-type geometry (linked with Jordan frame GR) dual to it.

Although the wormhole picture is not simpler than its conformal black hole one, it is more viable because these geometrical objects (Jordan frame wormholes) are invariant respecting transformations (3.6-3.12) that can be interpreted as particular transformations of physical units. As noted by Dicke\cite{dk}, these transformations should not influence the physics if the theory is correct. The Einstein frame Schwarzschild black hole, for his part, does not possess this invariance. More discussion on this point will be given in section VI.

\section{Geometrical duality in cosmology}

Other illustrations to the notion of geometrical duality come from cosmology. In the Einstein frame the FRW line element for flat space can be written as:

\begin{equation}
d\hat s^2=-dt^2+\hat a(t)^2(dr^2+r^2d\Omega^2),
\end{equation}
where $\hat a(t)$ is the EF scale factor.  Suppose the universe is filled with a perfect-fluid-type matter with the barotropic equation of state (in the EF): $\hat p=(\gamma-1)\hat \mu$, $0 < \gamma < 2$. Taking into account the line element (5.1) and the barotropic equation of state, the field equation (1.6) can be simplified to the following equation for determining the EF scale factor:

\begin{equation}
(\frac{\dot {\hat a}}{\hat a})^2=\frac{8\pi}{3} \frac{(C_2)^2}{\hat a^{3\gamma}},
\end{equation}
while, after integrating eq.(1.7) once, we obtain for the EF scalar:

\begin{equation}
\dot {\hat \phi}=\frac{C_1}{\hat a^3},
\end{equation}  
where $C_1$ and $C_2$ are arbitrary integration constants. The solution to eq.(5.2) is found to be:

\begin{equation}
\hat a(t)=(A)^{\frac{2}{3\gamma}} t^{\frac{2}{3\gamma}},
\end{equation}
where $A\equiv \sqrt{6\pi}\gamma C_2$. Integrating eq.(5.3) gives: 

\begin{equation}
{\hat \phi}^{\pm}(t)=\hat \phi_0 \mp B t^{1-\frac{2}{\gamma}},
\end{equation}
where $B\equiv \frac{\gamma C_1}{(2-\gamma)A^{\frac{2}{\gamma}}}$. 

The JF scale factor $a^{\pm}(t)=\hat a(t) \exp[-\frac{1}{2}\hat \phi^\pm (t)]$ is given by the following expression:

\begin{equation}
a^{\pm}(t)=\frac{A^\frac{2}{3\gamma}}{\sqrt{\phi_0}} t^\frac{2}{3\gamma} \exp[\pm \frac{B}{2} t^{1-\frac{2}{\gamma}}]. 
\end{equation}

The proper time $t$ in the EF and $\tau$ in the JF are related through:

\begin{equation}
(\tau-\tau_0)^{\pm}=\frac{1}{\sqrt{\phi_0}}\int \exp[\pm\frac{B}{2} t^{1-\frac{2}{\gamma}}] dt.
\end{equation}

For big $t$ ($t\rightarrow +\infty$) this gives $(\tau-\tau_0)^{\pm}\sim t$. Then $t\rightarrow +\infty$ implies $\tau\rightarrow +\infty$ for both '+' and '-' branches of our solution, given by the choice of the '+' and '-' signs in eq.(5.5).

For $t\rightarrow 0$, the r.h.s. of eq.(5.7) can be transformed into:

\begin{equation}
-\frac{\gamma}{\sqrt{\phi_0}}(2-\gamma)\int \frac{\exp[\pm Dx]}{x^\frac{2}{2-\gamma}} dx,
\end{equation}
where we have defined $x\equiv t^{1-\frac{2}{\gamma}}$ and $D\equiv \frac{\gamma C_1}{2(\sqrt{6\pi}\gamma C_2)^\frac{\gamma}{2}(2-\gamma)}$. If we take the '-' sign in the exponent under integral (5.8) then, for $t\rightarrow 0$ ($x\rightarrow\infty$), $\tau\rightarrow\tau_0$. If we take the '+' sign, for his part, integral (5.8) diverges for $t\rightarrow 0$ so $\tau\rightarrow -\infty$ in this last case.

In the '-' branch of our solution the evolution of the universe in the Jordan frame is basically the same as in the Einstein frame. The flat FRW perfect-fluid-filled universe evolves from a cosmological singularity at the beginning of time $t=0$ ($\tau=\tau_0$ in the JF) into an infinite size universe at the infinite future $t=+\infty$ ($\tau=+\infty$ in the JF). It is the usual picture in canonical general relativity where the cosmological singularity is unavoidable.

However, in the '+' branch of the solution the JF flat FRW perfect-fluid-filled universe evolves from an infinite size at the infinite past ($\tau=-\infty$) into an infinite size at the infinite future ($\tau=+\infty$) through a bounce at $t^*=[\frac{3}{4} \frac{\gamma C_1}{(\sqrt{6\pi}\gamma C_2)^{2\gamma}}]^{\frac {\gamma}{2-\gamma}}$ where it reaches its minimum size $a^*=\frac{1}{\sqrt{\phi_0}}[\sqrt{\frac{3}{32\pi}}\frac{C_1}{C_2} e]^{\frac{2}{3(2-\gamma)}}$. Then the Jordan frame universe is free of the cosmological singularity unlike the Einstein frame universe where the cosmological singularity is unavoidable. The more general case of arbitrary $\omega > -\frac{3}{2}$ is studied in \cite{qbc}.

If we model the JF scalar field $\phi$ as a perfect fluid then, in the Jordan frame its 'true' energy density (as measured by a comoving observer) will be given by the following expression:

\begin{equation}
\mu^\phi_\pm=\frac{(\omega+\frac{3}{2}) (C_1 \phi_0)^2}{16\pi A^\frac{4}{\gamma}  t^\frac{4}{\gamma}} \exp[\mp 2B t^{1-\frac{2}{\gamma}}],
\end{equation}
while the 'effective' energy density of $\phi$ as seen by a cosmological observer is given by:

\begin{equation}
\mu^\phi_{eff,\pm}=\frac{(\omega+3) (C_1 \phi_0)^2}{16\pi A^\frac{4}{\gamma}  t^\frac{4}{\gamma}} \exp[\mp 2B t^{1-\frac{2}{\gamma}}](1-\frac{4A^\frac{2}{\gamma}t^{\frac{2}{\gamma}-1}}{(\omega+3)\gamma C_1}).
\end{equation}

In the '+' branch of the JF solution both $\mu^\phi$ and $\mu^\phi_{eff}$ are finite for all times. In this case the 'true' energy density (equation (5.9)) evolves from zero value at $t=0$ ($\tau=-\infty$) into a zero value at the infinite future ($\tau=+\infty$), through a maximum (finite) value at some intermediate time. It is positive definite for all times. This means that the scalar matter is non-exhotic and non-singular for all times. The 'effective' scalar field energy density (5.10) evolves from zero at $t=0$ ($\tau=-\infty$) into zero at $t^*=[(\omega+1)(2-\gamma)]^\frac{\gamma}{2-\gamma}$ through a maximum (finite value) at some prior time. In this range of times $\mu^\phi_{eff}$ is positive definite. Then it further evolves from a zero value at $t^*$ into a zero value at $t=+\infty$ ($\tau=+\infty$) through a maximum absolute value at some time $t^*<t<+\infty$. In this range of times $\mu^\phi_{eff}$ is negative definite.

For the perfect fluid barotropic ordinary matter we found that the energy density in the 'plus' branch of the Jordan frame solution is given by:

\begin{equation}
\mu=(\frac{C_2}{A})^2 \frac{\phi_0}{t^2} \exp[-B t^{1-\frac{2}{\gamma}}].
\end{equation}

It evolves from zero at $t=0$ ($\tau=-\infty$)into zero density at $t=+\infty$ ($\tau=+\infty$) through a maximum value $\mu^*= e^{-2}[A^{6-\gamma}(\frac{2(2-\gamma)}{\gamma C_1})^{2\gamma}]^\frac{1}{2-\gamma}$ at $t^*=(\frac{\gamma C_1}{2(2-\gamma)})^\frac{\gamma}{2-\gamma}\frac{1}{A^\frac{2}{2-\gamma}}$, i.e. it is bounded for all times. This means that the energy density as measured by a comoving observer is never singular (it is true for the sum of (5.9) and (5.11) as well as for the sum of (5.10) and (5.11)).

\section{The Jordan frame or the Einstein frame afterall?}

In this section we are going to discuss about the physical implications of the viewpoint developed in the present paper. Our proposal is based on the postulate that the different conformal formulations of general relativity are physically equivalent. Among these conformal representations of GR the Jordan frame and the Einstein frame formulations are distinguished. 

For the purpose of the present discussion we shall take the static, spherically symmetric solution presented in section IV. In the Einstein frame for $\omega=-\frac{3}{2}$ this is the typical Schwarzschild black hole solution. A time-like singularity at the origin of the Schwarzschild radial coordinate is enclosed by an event horizon at $r=2m$. To a distant observer, an observer falling into the black hole asymptotically approaches the event horizon but he never crosses the surface $r=2m$. The same is true for a distant observer in the Jordan frame because spacetime coincidences are not affected by a aconformal transformation of the metric. This means that to a distant observer, the black hole never forms neither in the EF nor in the JF. The situation is dramatically different for an observer falling into the black hole (in the JF we have a wormhole instead of a black hole). In the Einstein frame he crosses the event horizon and inevitably hits the singularity at $r=0$ in a finite proper time. Unlike this, in the Jordan frame this observer never sees any singularity (he never crosses the surface $r=2m$). For $\omega>-\frac{3}{2}$, in the Einstein frame to a distant observer, an observer falling into the singular point $r=\frac{2m}{p}$ will reach it in a finite time. In this case the singularity at $\frac{2m}{p}$ is naked so it is seen by the distant observer. The same is true for a distant observer in the Jordan frame. He finds  that the observer falling into the surface $r=\frac{2m}{p}$ (the begining of the JF proper radial coordinate $\rho=0$) will reach it in a finite time. However, in this case, the surface $r=\frac{2m}{p}$ is non-singular for $q\geq2-p$ ($0<p\leq1$) because the curvature scalar $R=-\frac{6m^2 q^2}{p^2 r^4}(1-\frac{2m}{pr})^{p+q-2}$ is finite at $\frac{2m}{p}$. In the Einstein frame, the observer falling into the singular point $r=\frac{2m}{p}$ hits the singularity in a finite proper time. In the Jordan frame the falling observer never meets any singularity. Moreover, it takes an infinite proper time to the falling observer to reach the surface $r=\frac{2m}{p}$.

Summing up. The physics as seen by a distant observer is the same in the Einstein and in the Jordan frames since spacetime coincidences are unchanged under the conformal rescaling of the metric. Unlike this, the physics seems dramatically different to the falling observer in the Einstein frame. He hits the singularity in a finite proper time. In the Jordan frame the falling observer never meets any singularity. It is very striking because, according to our proposal, both situations are physically equivalent (they are equally consistent with the observational evidence). However, the falling observer is part of the physical reality and the physical reality is unique(it is a prime postulate in physics). 

We can not pretend to give a final answer to this paradoxical situation since, we feel it is a very deep question. We shall however conjecture on this subject. Two explanations to this striking situation are possible. The first one is based on the fact that Einstein's theory is a classical theory of spacetime and near of the singular point we need a quantum theory (this theory has not been well stablished at present). When a viable quantum gravity theory will be worked out it may be that this singularity is removed. In the Jordan frame no singularity occurs (for $q\geq2$) and, consequently, we do not need of any quantum theory for describing gravitation. This explanation is in agreement with a point of view developed in reference \cite{shojai}. According to this viewpoint, to bring in the quantum effects into the classical gravity theory one needs to make only a conformal transformation. If we start with Einstein's classical theory of gravitation then we can set in the quantum effects of matter by simply making a conformal transformation into, say, the Jordan frame. In this sense the Jordan frame formulation of general relativity already contains the quantum effects (Jordan frame GR represents a unified description of both gravity and the quantum effects of matter). 

The second possibillity is more radical and has been already outlined above in this paper. The Einstein frame formulation is not invariant under the particular transformations of the units of time, length and mass studied in section III. It is very striking since the physical laws should be invariant under the transformations of units. Unlike this The Jordan frame formulation of general relativity is invariant in respect to these transformations. This means that the picture without singularity is more viable than the one with them, i.e. spacetime singularities are not physical. These are fictitious entities due to a wrong choice of the formulation of the theory.

We recall that these are just conjectures and we hope to discuss further on this point in future works.

\begin{center}
{\bf AKNOWLEDGEMENT}
\end{center}

We thank A. G. Agnese and D. Wands for helpful comments. We also aknowlege the unknown referees for recommendations and criticism and MES of Cuba for financing.

\end{document}